\documentstyle[aps,floats,preprint,epsfig]{revtex}
\tightenlines

\begin{document}
\draft
\preprint{\vbox{
\hbox{hep-ph/0108268}
\hbox{Aug 31, 2001}}}
\title{Signals for Non-Commutative QED in $e \gamma$ and $\gamma \gamma$ 
Collisions}
\author{Stephen Godfrey}
\address{ Ottawa-Carleton Institute for Physics \\
Department of Physics, Carleton University, Ottawa, Canada K1S 5B6}
\author{M.A. Doncheski}
\address{Department of Physics, Pennsylvania State University, \\
Mont Alto, PA 17237 USA}

\maketitle

\begin{abstract}
We study the effects of non-commutative QED (NCQED) in fermion pair 
production, $\gamma + \gamma \rightarrow f + \bar{f}$ and Compton scattering, 
$e + \gamma \rightarrow e + \gamma$.  Non-commutative geometries appear 
naturally in the context of string/M-theory and gives rise to 3- and 4-point 
photon vertices and to momentum dependent phase factors in QED vertices which 
will have observable effects in high energy collisions.  We consider 
$e^+ e^-$ colliders with energies appropriate to the TeV Linear Collider 
proposals and the multi-TeV CLIC project operating in $\gamma \gamma$ and 
$e\gamma$ modes.  Non-commutative scales roughly equal to the center of mass 
energy of the $e^+e^-$ collider can be probed, with the exact value depending 
on the model parameters and experimental factors.  However, we found that the 
Compton process is sensitive to $\Lambda_{NC}$ values roughly twice as large 
as those accessible to the pair production process.
\end{abstract}

\section{Introduction}

Although string/M-theory is still developing, and the details of its 
connection to the Standard Model are still unclear, numerous ideas from 
string/M-theory have affected the phenomenology of particle physics.  The 
latest of these ideas is non-commutative quantum field theory (NCQFT) 
\cite{Douglas,ncqft}.  NCQFT arises through the quantization of strings by 
describing low energy excitations of D-branes in background EM fields.  NCQFT 
generalizes our notion of space-time, replacing the usual, commuting, 
space-time coordinates with non-commuting space-time operators.  This is 
similar to the replacement of the commuting position and momentum coordinates 
of classical physics with the non-commuting position and momentum operators 
of quantum mechanics. Significant, testable differences exist between QFT with 
commuting space-time coordinates and NCQFT.  This article is an attempt to 
probe those changes.

At this time, the details of a general NCQFT model to compare to the Standard 
Model are just emerging \cite{ncsm}.  However, a non-commuting replacement of 
quantum electrodynamics, NCQED, does exist and can be studied.  NCQED modifies 
QED, with the addition of a non-Lorentz invariant, momentum dependent phase 
factor to the normal $ee\gamma$ vertex, along with the addition of cubic 
($\gamma \gamma \gamma$) and quartic ($\gamma \gamma \gamma \gamma$) coupling, 
also, with non-Lorentz invariant momentum dependent phase factors.  The 
Feynman rules for NCQED are given in \cite{frncqed,hpr}, and will not be 
repeated here.  Although the momentum dependent phase factors and higher 
dimensional operators in the Lagrangian (leading to additional couplings) 
arise naturally in NCQFT, the modifications, although similar, will in 
general, take on a different form than  those presented here for NCQED.  We 
will see that the modifications of NCQFT to QED can be probed in 
$\gamma\gamma \to f\bar{f}$ and $e\gamma \to e\gamma$ collisions. 

The essential idea of NCQFT is that in the non-commuting space time the 
conventional coordinates are represented by operators which no longer commute:
\begin{equation}
[\hat{X}_\mu, \hat{X}_\nu] = i\theta_{\mu\nu} \equiv {i\over 
{\Lambda_{NC}^2}} C_{\mu\nu}
\end{equation}
Here we adopt the Hewett-Petriello-Rizzo parametrization \cite{hpr} where the 
overall scale, $\Lambda_{NC}$, characterizes the threshold where 
non-commutative (NC) effects become relevant and $C_{\mu\nu}$ is a real 
antisymmetric matrix whose dimensionless elements are presumably of order 
unity.  One might expect the scale $\Lambda_{NC}$ to be of order the Planck 
scale.  However, given the possibility of large extra dimensions \cite{add,rs} 
where gravity becomes strong at scales of order a TeV, it is possible that NC 
effects could set in at a TeV.  We therefore consider the possibility that 
$\Lambda_{NC}$ may lie not too far above the TeV scale.

The $C$ matrix is not a tensor since its elements are identical in all 
reference frames resulting in the violation of Lorentz invariance.  The 
$C_{\mu\nu}$ matrix is related to the Maxwell field strength tensor 
$F_{\mu\nu}$ since NCQFT arises from string theory in the presence of 
background electromagnetic fields.  Hence, $C$ can be parameterized, following 
the notation of \cite{JoA}, as
\begin{equation}
C_{\mu \nu} = \left(
\begin{array}{cccc}
0 & C_{01} & C_{02} & C_{03} \\
-C_{01} & 0 & C_{12} & -C_{13} \\
-C_{02} & -C_{12} & 0 & C_{23} \\
-C_{03} & C_{13} & -C_{23} & 0 \\
\end{array}
\right)
\end{equation}
where $\sum_i |C_{0i}|^2 = 1$.  Thus, the $C_{0i}$ are related to space-time 
NC and are defined by the direction of the background {\bf E}-field.  
Furthermore, the $C_{0i}$ can be parameterized as 
\begin{eqnarray}
C_{01} & = & \sin \alpha \cos \beta \nonumber \\
C_{02} & = & \sin \alpha \sin \beta \nonumber \\
C_{03} & = & \cos \alpha.
\end{eqnarray}
$\beta$ defines the origin of the $\phi$ axis which we set to $\beta = \pi/2$ 
and $\alpha$ is the angle of the background {\bf E}-field relative to the 
$z$-axis.  Likewise, the $C_{ij}$ are related to the space-space 
non-commutativeness and are defined by the direction of the background 
{\bf B}-field.  They can be parameterized as 
\begin{eqnarray}
C_{12} & = & \cos \gamma \nonumber \\
C_{13} & = & \sin \gamma \sin \beta \nonumber \\
C_{23} & = & -\sin \gamma \cos \beta.
\end{eqnarray}

NCQFT can be cast in the form of conventional commuting QFT through the 
application of Weyl-Moyal correspondance \cite{ihab}.  The details of this 
derivation is given by Ref. \cite{hpr}.  The net result is that the QED 
vertices pick up phase factors dependent on the momenta flowing through them 
and three and four point photon vertices are now present.  These NCQED 
modifications are what is being tested in collider tests of NCQED.  In 
addition,  covariant derivatives can only be constructed for (fermion) fields 
of charge $0, \; \pm 1$ so we restrict our analysis to processes involving 
only charged leptons.  The Feynman rules for NCQED are given in 
Ref.~\cite{frncqed,hpr}.

NCQED is beginning to attract theoretical and phenomenological interest 
\cite{hpr,ncpheno,mathews,bghh}.  Hewett, Petriello and Rizzo~\cite{hpr} have 
performed a series of phenomenological studies of NCQED at high energy, 
linear, $e^+ e^-$ colliders.  They analyzed diphoton production 
($e^+ + e^- \to \gamma + \gamma$), Bhabha scattering 
($e^+ + e^- \to e^+ + e^-$) and Moller scattering 
($e^- + e^- \to e^- + e^-$).  There are striking differences between QED and 
NCQED for all three processes; most interesting is significant structure in 
the $\phi$ angular distribution.

Mathews \cite{mathews}  and Baek, Ghosh, He and Hwang~\cite{bghh} have also 
studied NCQED at high energy $e^+ e^-$ linear colliders.  In the former case 
Mathews studied high energy Compton scattering while Baek {\it et al.,} 
studied fermion pair production in $\gamma + \gamma \to e^+ + e^-$.  In both 
cases the initial state photons are due to backscattering of laser photons off 
the electron and positron beams.  As is well known, this produces a high 
luminosity, high energy photon beam, effectively converting an $e^+ e^-$ 
collider to an $e \gamma$ or $\gamma \gamma$ collider \cite{backlaser}.  
Independently of the aforementioned studies we studied Compton scattering and 
lepton pair production.  In our study we studied the angular distributions, in 
contrast to the work of Mathews \cite{mathews} and Baek {\it et al.,} 
\cite{bghh} whose analysis is based on the total cross section and which do 
not use the additional information inherent in the angular distributions.  We 
find that the analysis based on angular distributions leads to exclusion 
limits on the NCQED scale of order 100~GeV or more greater than those obtained 
by simply measuring the total cross section. In addition we also studied the 
effect on sensitivity of including systematic errors in addition to 
statistical errors.  There are a number of other differences between our work 
and that of these authors.  In the first case, Mathews seems to have 
calculated the NC phase appearing in the cross section in the $e\gamma$ center 
of mass.  This is an inherently Lorentz violating quantity and we believe that 
it should be to calculated in the lab frame.  We therefore disagree with his 
approach.  In the case of the work of Baek {\it et al.} we decided on 
different kinematic cuts which we feel to be more realistic.  Ultimately, the 
best approach will be decided by experimentalists, based on detailed detector 
simulations.  To this end, it may be of some use to see the tradeoffs inherent 
in different approaches.

In the following sections we will examine the effects of NCQED in 
$\gamma\gamma\to f\bar{f}$ and Compton scattering, $e\gamma \to e\gamma$.  In 
the case of Compton scattering NCQED leads to an oscillatory azimuthal 
dependence due to the preferred direction in the laboratory frame defined by 
the $C$ matrix.  As will be discussed in detail later, we find that the 
Compton scattering process yields significantly higher exclusion limits than 
the pair production process, despite lower statistics.  

Before proceeding we reiterate that the $C_{\mu \nu}$ matrix is not Lorentz 
invariant and the vectors $C_{i0}$ and $C_{ij}$ point in specific directions 
which are the same in all reference frames.  In our analysis we define the 
$z$-axis to correspond to the direction of the incoming particles in the lab 
frame.  If the experiment were to be repeated at a different location, the 
co-ordinates will be in general be different.  In fact, as the earth rotates 
and revolves around the Sun, the co-ordinate system also rotates.  Hence,  it 
is important that the local co-ordinates be converted to a common frame such 
as a slowly varying astronomical co-ordinate system so that all measurements 
are made with respect to a common frame.    More germaine to our specific 
examples is that one must calculate the cross sections in the lab frame not 
the centre of mass frame of either the initial $\gamma\gamma$ or $e\gamma$ 
beams since each event will have a different momentum fraction of the initial 
electron beams and hence different boosts between the lab and center of mass 
frames. 

\section{Calculations and Results}

We begin by discussing the common points of our two analyses.  We will present 
details and results from the pair production and Compton scattering prosesses 
in separate subsections below.

In both cases, we consider linear $e^+ e^-$ colliders operating at 
$\sqrt{s} = 0.5$ and 0.8~TeV appropriate to the TESLA proposal, \cite{tesla}
$\sqrt{s} = 0.5$, 1.0 and 1.5~TeV as advocated by the NLC proponents 
\cite{nlc}, and $\sqrt{s} = 3.0$, 5.0 and 8.0~TeV being considered in CLIC 
studies \cite{clic}.  In order to estimate event rates, we assume an 
integrated luminosity of $L = 500$~fb$^{-1}$ for all cases.  We impose 
acceptance cuts on the final state particles of $10^o \leq \theta \leq 170^o$ 
and $p_{_T} > 10 \; GeV$.  

As noted above, we take $\beta = \pi/2$. Therefore, in the pair production 
case, where only space-time NC enters, only the parameter $\alpha$ remains in 
addition to $\Lambda_{NC}$.  We consider three specific cases, $\alpha = 0$, 
$\pi/4$ and $\pi/2$, and report limits on $\Lambda_{NC}$ for each of these 
values.  In the Compton scattering case, both space-space and space-time NC 
enter, leaving the two parameters, $\alpha$ and $\gamma$ in addition to 
$\Lambda_{NC}$.  We  examine the two values $\gamma=0$ and $\gamma = \pi/2$, 
and for each value of $\gamma$ give exclusion limits for $\alpha = 0$, $\pi/4$ 
and $\pi/2$.

In order to quantify the sensitivity to NCQED, we calculate the $\chi^2$ for 
the deviations between NCQED and the SM for a range  of parameter values.  We 
start by calculating statistical errors based on an integrated luminosity of 
500~fb$^{-1}$.  We assume that the statistical errors are gaussian, which 
given the large event rates, is certainly valid.  We consider two 
possibilities for systematic errors.  In the first case we do not include 
systematic errors while in the second case we obtain limits by combining a 2\% 
systematic error combined in quadrature with the statistical errors; 
$\delta =\sqrt{\delta^2_{stat} + \delta^2_{sys}}$.  The 2\% systematic error 
is a very conservative estimate of systematic errors, for example the TESLA 
TDR calls for only a 1\% systematic error.  Our exclusions limits including 
systematic errors should therefore be considered conservative estimates of 
those thought to be eventually achievable.  Next, we calculate total cross 
sections, and $\cos \theta$ and $\phi$ angular distributions in both QED and 
NCQED.  We bin the angular distributions into 20 bins in $\cos \theta$ and 
$\phi$.  Finally, we calculate the $\chi^2$ for the different observables, 
$\cal O$, using:
\begin{equation}
\chi^2_{\cal O} (\Lambda) 
=   \sum_i
\left( { \frac{ {\cal O}_i (\Lambda) - {\cal O}_i^{QED} }
{\delta {\cal O}_i } }\right) ^2 
\end{equation}
where $\cal O$ represents the observable under consideration and the sum is 
over the bins of the angular distributions.  $\chi^2 = 4$ represents a 
95\% C.L. deviation from QED, which we'll define as the sensitivity limit.

\subsection{Pair Production}

For the pair production process, Figure 1 shows the Feynman diagrams that 
contribute.  Note the presence of the novel s-channel contribution from the 
presence in NCQED of the $3\gamma$ self-coupling.  The differential cross 
section for this process is given by:
\begin{equation}
{{d\sigma(\gamma\gamma\to f\bar{f})}\over{d\cos\theta \; d\phi}} = {{\alpha^2}\over{2s}}
\left\{ \frac{\hat{u}}{\hat{t}} + \frac{\hat{t}}{\hat{u}} - 
4 \frac{\hat{t}^2 + \hat{u}^2}{\hat{s}^2} \sin^2 
\left(\frac{k_1 \cdot \theta \cdot k_2}{2} \right) \right\}.
\end{equation}
The first two terms in the expression are the standard QED contributions, 
while the last term is due to the Feynman diagram with the cubic 
$\gamma \gamma \gamma$ coupling.  The phase factor, 
$\sin^2 \left(\frac{\mbox{$k_1 \cdot \theta \cdot k_2$}}{\mbox{$2$}} \right)$ 
only appears in this new term.  $p_1$ and $p_2$ are the momentum of the 
electron and positron, respectively, while $k_1$ and $k_2$ are the momenta of 
the incoming photons.  $\hat{s}$, $\hat{t}$ and $\hat{u}$ are the usual 
Mandelstam variables $\hat{s} = (k_1 + k_2)^2$, $\hat{t} = (k_1 - p_1)^2$ and 
$\hat{u} = (k_1 - p_2)^2$.  $k_1$ and $k_2$ are given by 
\begin{equation}
k_1= {{x_1\sqrt{s}}\over{2}} (1, 0, 0, 1) \quad \hbox{and} \quad 
k_2= {{x_2\sqrt{s}}\over{2}} (1, 0, 0, -1)
\end{equation}
where $x_1$ and $x_2$ are the momentum fractions of the two photons and the 
4-vectors follow the convention of $k=(E, k_x, k_y, k_z)$.  With this 
definition the bilinear product in eqn. 6 simplifies to 
\begin{equation}
\frac{1}{2} k_1 \cdot \theta \cdot k_2 = \frac{\hat{s}}{4 \Lambda_{NC}^2} 
C_{03}.
\end{equation}
The expression for the cross section is not Lorentz invariant due to the 
presence of the phase factor.  Note that only space-time non-commutativity 
contributes and there is no $\phi$ dependence in this case.  In the limit 
$\Lambda_{NC}\to \infty$ the angle goes to zero and the SM is recovered.  
Given that we've chosen $\beta = \pi/2$, $C_{03} = \cos \alpha$, and the phase 
factor is identically zero for $\alpha = \pi/2$.  Thus, for $\alpha = \pi/2$, 
the NCQED and QED calculations should be identical, and {\bf no} limits on 
$\Lambda_{NC}$ are possible for $\alpha = \pi/2$.  

Fig.~2 shows the cross section for $\gamma\gamma\to e^+e^-$ vs. $\Lambda_{NC}$ 
for QED and NCQED with $\alpha = 0$ and $\pi/4$, for a $\sqrt{s} = 0.5$~TeV 
$e^+ e^-$ collider operating in $\gamma \gamma$ mode.  The event rate is high 
with statistics that can exclude NCQED to a fairly high value of 
$\Lambda_{NC}$.  Note that the QED (solid) curve is actually a central QED 
value with $\pm 1 \sigma$ bands (assuming 500~fb$^{-1}$ of integrated 
luminosity).  Fig.~3 shows the $\cos\theta$ angular distribution, 
$d\sigma/d\cos \theta$ for QED and NCQED with $\alpha = 0$, and 
$\sqrt{s} = 500 \; GeV$ and $\Lambda_{NC} = 300 \; GeV$.

We calculated the significance of deviations from the SM using the total cross 
section and by binning the angular distribution.  We found that the 
$\cos \theta$ distribution consistently gives the highest exclusion limits on 
$\Lambda_{NC}$, regardless of $\sqrt{s}$ and $\alpha$ (as long as 
$\alpha \neq \pi/2$, where, again, no limits are possible). 

The exclusion limits based on lepton pair production in $\gamma\gamma$ 
collisions and assuming an integrated luminosity of $L = 500 fb^{-1}$ are 
summarized in Table I for $\alpha = 0$ and $\pi/4$.  These are based on the 
angular distribution which, as already noted, gives the highest limits.  These 
limits could be improved by including three lepton generations in the final 
state and assuming some value for the lepton detection efficiency.

We also considered the limits on the NC-scale that could be obtained in 
$e^+e^-$ collisions using Weisz\"acker-Williams photons.  Assuming 
500~fb$^{-1}$ of integrated luminosity and no systematic errors for 
$\sqrt{s}=500$~GeV and 5~TeV, $\Lambda_{NC}$ can be probed to 175~GeV and 
370~GeV respectively for $\alpha=0$.  These limits are pretty much irrelevant 
compared to the limits that can be obtained in the more direct processes of 
Bhabba scattering and $e^+e^-\to \gamma\gamma$ in high energy $e^+e^-$ 
collisions \cite{hpr}.

\subsection{Compton scattering}

For the Compton scattering process, Figure 4 shows the Feynman diagrams that 
contribute. We find:
\begin{equation}
{{d\sigma(e^-\gamma\to e^-\gamma)}\over{d\cos\theta \; d\phi}} = {{\alpha^2}\over{2s}}
\left\{ -\frac{\hat{u}}{\hat{s}} - \frac{\hat{s}}{\hat{u}} + 
4 \frac{\hat{s}^2 + \hat{u}^2}{\hat{t}^2} \sin^2 
\left(\frac{k_1 \cdot \theta \cdot k_2}{2} \right) \right\}.
\end{equation}
The first two terms in the expression are the standard, QED contribution, 
while the last term is due to the Feynman diagram with the cubic 
$\gamma \gamma \gamma$ coupling.  As before, the phase factor only appears in 
this new term.

Here, $p_1$ and $k_1$ are the momenta of the initial state electron and 
photon, respectively, while $p_2$ and $k_2$ are the momenta of the final state 
electron and photon, respectively.  $\hat{s}$, $\hat{t}$ and $\hat{u}$ are the 
usual Mandelstam variables $\hat{s} = (p_1 + k_1)^2$, 
$\hat{t} = (p_1 - p_2)^2$ and $\hat{u} = (p_1 - k_2)^2$.  Choosing 
$k_1 = x \frac{\mbox{$\sqrt{s}$}}{\mbox{$2$}} (1,0,0,-1)$ and 
$k_2 = k(1,\sin \theta \cos \phi, \sin \theta \sin \phi, \cos \theta)$, the 
phase factor can be evaluated analytically:
\begin{equation}
\frac{1}{2} k_1 \cdot \theta \cdot k_2 = \frac{x k \sqrt{s}}{4 \Lambda_{NC}^2} 
[ (C_{01} - C_{13}) \sin \theta \cos \phi + 
(C_{02} + C_{23}) \sin \theta \sin \phi + C_{03}(1 + \cos \theta) ].
\end{equation}
where $x$ is the momentum fraction of the incident photon, $k$ is the 
magnitude of the 3-momentum of the final state photon, and $\theta$ and $\phi$ 
are the lab frame angles of the final state photon.  Note that there is no 
$C_{12}$ term appearing in the above expression since defining the $z$-axis 
along the beam direction results in no {\bf B} field in the $C_{12}$ 
direction.  It is clear that this phase factor includes both space-space and 
space-time NC parts, so this process probes $\gamma$, in addition  to $\alpha$ 
and $\beta$.  We will again choose $\beta = \pi/2$, leaving us two free 
parameters in adddition to $\Lambda_{NC}$.  In this case the phase factor 
simplifies to
\begin{equation}
\frac{1}{2} k_1 \cdot \theta \cdot k_2 = \frac{x k \sqrt{s}}{4 \Lambda_{NC}^2} 
[ - \sin \gamma \sin \theta \cos \phi + 
\sin \alpha \sin \theta \sin \phi + \cos \alpha (1 + \cos \theta) ].
\end{equation}
We remind the reader that $\alpha=0$ corresponds to {\bf E} parallel to the 
$z$-axis and $\alpha=\pi/2$ corresponds to {\bf E} perpendicular to the 
$z$-axis.  Because Compton scattering is sensitive to both $\gamma$ and 
$\alpha$, it is complimentary to the pair production process studied above.

After analyzing our results, the total cross section consistently gives the 
weakest exclusion limits on $\Lambda_{NC}$.  For $\gamma = 0$, the 
$\cos \theta$ distribution gives the strongest exclusion limits when 
$\alpha = 0$ or $\pi/2$, while the $\phi$ distribution gives the highest 
exclusion limits when $\alpha = \pi/4$.  For $\gamma = \pi/2$, the $\phi$ 
distribution gives the highest exclusion limits when $\alpha = 0$, while the 
$\cos \theta$ distribution gives the highest exclusion limits when 
$\alpha = \pi/4$ and $\pi/2$.  When including a 2\% systematic uncertainty, 
the $\phi$ distribution becomes more important: for $\gamma = 0$ the $\phi$ 
distribution gives the highest exclusion limits for $\alpha = \pi/4$ or 
$\pi/2$, while for $\gamma = \pi/2$, the $\phi$ distribution gives the 
highest exclusion limits for all values of $\alpha$ tested.

\subsubsection{$\gamma = 0$ exclusion limits}

Fig.~5 shows the cross section $\sigma$ vs. $\Lambda_{NC}$ for QED and NCQED 
with $\alpha = 0$, $\pi/4$ and $\pi/2$, for a $\sqrt{s} = 0.5 \; TeV$ 
$e^+ e^-$ collider operating in $e \gamma$ mode.  The event rate is high, so 
there are enough statistics to probe NCQED up to a fairly high value of 
$\Lambda_{NC}$.  Again, the QED (solid) curve includes the central QED value 
and $\pm 1 \sigma$ bands (assuming 500~fb$^{-1}$ of integrated luminosity).  
Fig.~6a and Fig.~6b show the angular distributions, $d\sigma/d\cos \theta$ and 
$d\sigma/d\phi$, for QED and NCQED with $\alpha = \pi/2$, and 
$\sqrt{s}$ = $\Lambda_{NC}$ = $500 \; GeV$.  The error bars in Fig.~6 assume 
500~fb$^{-1}$ of integrated luminosity.

Note that there is no $\phi$ dependence for $\alpha=0$ since for this case 
both {\bf E} and {\bf B} are parallel to the beam direction.  In contrast, when
$\alpha=\pi/2$, {\bf E} is perpendicular to the beam direction which is 
reflected in the strong oscillatory behavior in the $\phi$ distribution. 

The exclusion limits obtainable from Compton scattering are summarized in 
Table II for $L = 500 fb^{-1}$.  Limits are given for the three values of 
$\alpha =0$, $\alpha = \pi/4$, and $\alpha = \pi/2$.  We give the highest 
limits obtained from the total cross section, $d\sigma/d\cos\theta$ or 
$d\sigma/d\phi$.  With no systematic errors the $\cos\theta$ distribution gave 
the best limits for $\alpha = 0$ and $\pi/2$, while the $\phi$ distribution 
gives the highest exclusion limits when $\alpha = \pi/4$.  When systematic 
errors are included the $\phi$ distribution gave the best limits except for 
the case $\alpha=\gamma=0$ where there is no $\phi$ dependence. 

\subsubsection{$\gamma = \pi/2$ exclusion limits}

Fig.~7 shows the cross section $\sigma$ vs. $\Lambda_{NC}$ for QED and NCQED 
with $\alpha = 0$, $\pi/4$ and $\pi/2$, for a $\sqrt{s} = 0.5 \; TeV$ 
$e^+ e^-$ collider operating in $e \gamma$ mode.  Again, the QED (solid) curve 
includes the central QED value and $\pm 1 \sigma$ bands (assuming 
500~fb$^{-1}$ of integrated luminosity).  Fig.~8a and Fig.~8b show the angular 
distributions, $d\sigma/d\cos \theta$ and $d\sigma/d\phi$, for QED and NCQED 
with $\alpha = \pi/2$, and $\sqrt{s}$ = $\Lambda_{NC}$ = $500 \; GeV$.  The 
error bars in Fig.~8 assume $500 fb^{-1}$ of integrated luminosity.  The 
exclusions limits for these cases are given in Table II.  With no systematic 
errors, when $\gamma = \pi/2$, the $\phi$ distribution gives the highest 
exclusion limits when $\alpha = 0$, while the $\cos \theta$ distribution gives 
the highest exclusion limits when $\alpha = \pi/4$ and $\pi/2$.  When 
including a 2\% systematic uncertainty, the $\phi$ distribution gives the 
highest exclusion limits for all values of $\alpha$ tested.

\section{Conclusions}

In conclusion, we found that lepton pair production and Compton scattering at 
high energy linear colliders are excellent processes to study non-commutative 
QED.  These processes compliment those studied by Hewett, Petriello and 
Rizzo~\cite{hpr}.

The pair production process is only sensitive to space-time NC and is 
therefore insensitive to $\gamma$.  As $\alpha$ increases towards $\pi/2$ the 
deviations from SM decrease towards zero, with $\alpha = \pi/2$ being 
identical to the SM.  On the other hand, the Compton scattering process is 
sensitive to both space-space and space-time NC as parametrized by $\gamma$ 
and $\alpha$.  On the whole, we found that the Compton scattering process is 
superior to lepton pair production in probing NCQED.  Despite significantly 
smaller statistics, the large modification of angular distributions 
(see Fig.~6 and 8) leads to higher exclusion limits, well in excess of the 
center of mass energy for all colliders considered.

After the completion of this work  M.~Chaichian {\it et al.} \cite{ncsm} 
presented a model for the NC SM.  The primary implication from  NCSM vs NCQED 
in the context of our calculations is the introduction of a $\gamma\gamma Z$ 
vertex.  Although this will alter details of our results we do not expect it 
to change our main conclusions. This is the philosophy followed by Hewett 
{\it et al.} \cite{hpr}.

\acknowledgments

The authors thank JoAnne Hewett and Tom Rizzo for many helpful conversations 
and communications.  This research was supported in part by the Natural 
Sciences and Engineering Research Council of Canada.  M.A.D. would like to 
thank the Physics Department at Carleton University where much of this work 
was performed.

\newpage
\begin{figure}
\begin{center}
\centerline{\epsfig{file=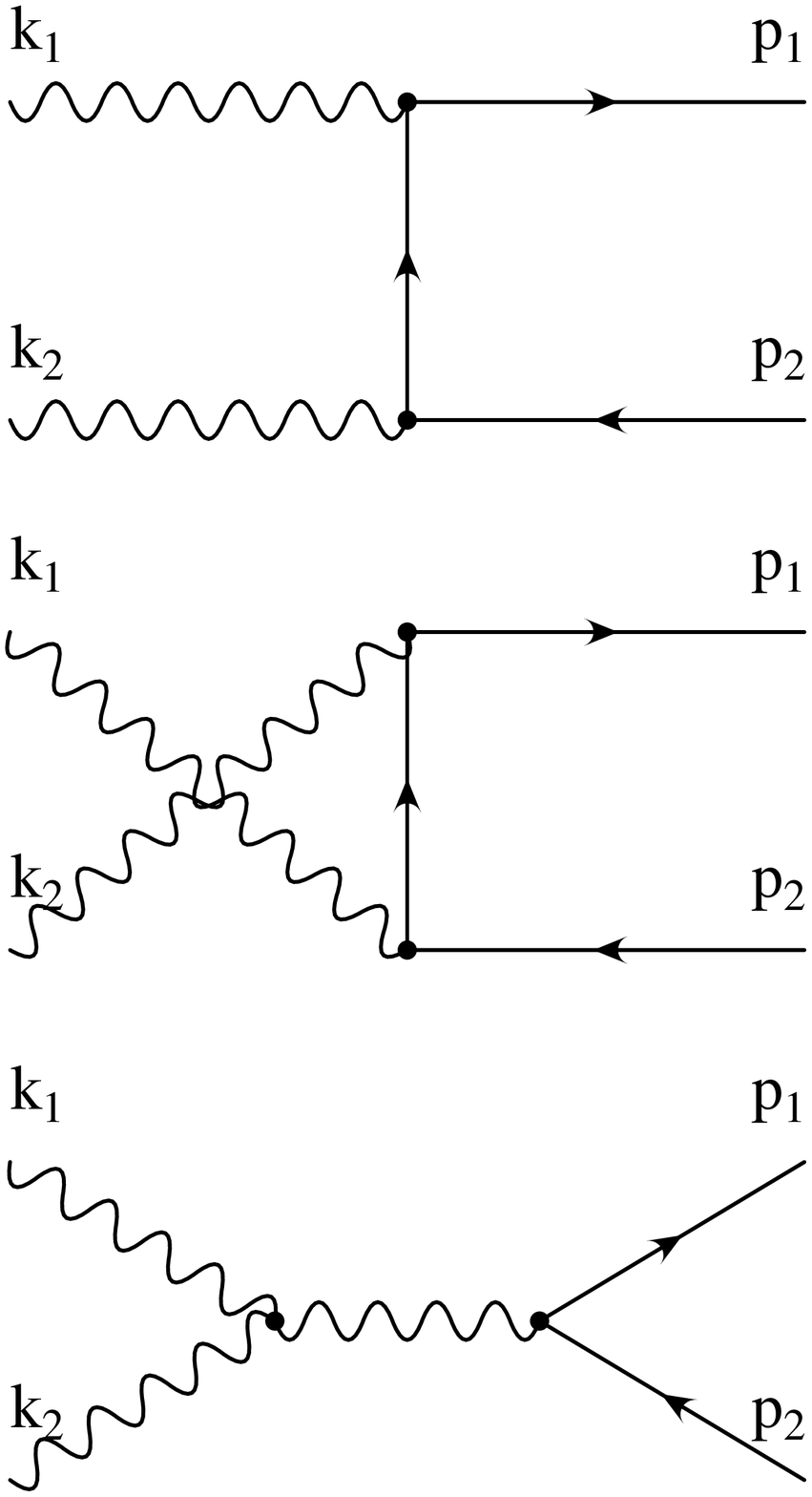,width=3.0in}}
\end{center}
\vspace{20pt}
\caption{The Feynman diagrams contributing to the process 
$\gamma \gamma \to e^+ e^-$.}
\label{Fig1}
\end{figure}

\newpage
\begin{figure}
\centerline{\epsfig{file=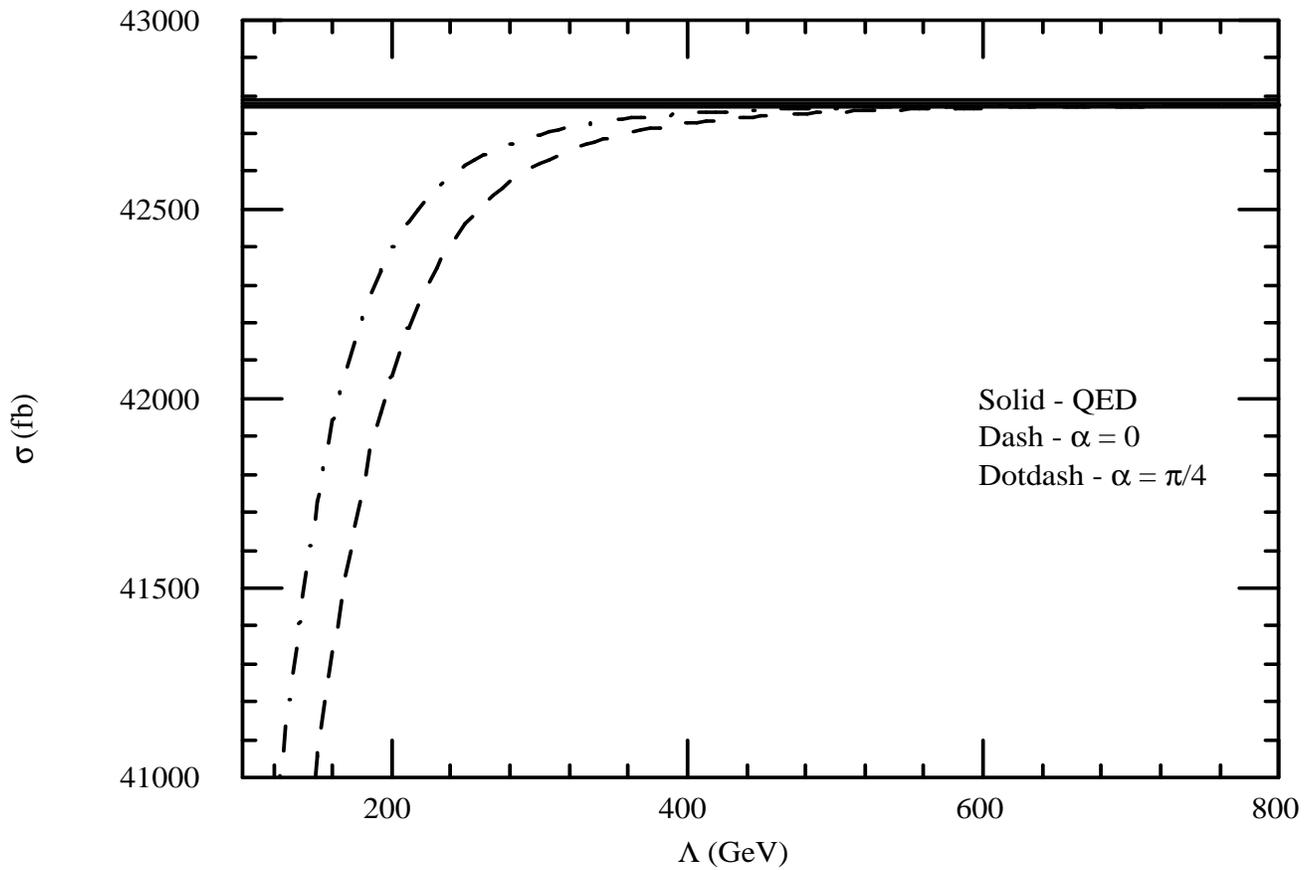,width=4.5in,angle=-90}}
\vspace{20pt}
\caption{$\sigma$ vs. $\Lambda_{NC}$ for the pair production process, 
$\sqrt{s} = 500$ GeV.  The solid line corresponds to the SM cross section 
$\pm$ 1 standard deviation (statistical) error.}
\label{Fig2}
\end{figure}

\newpage
\begin{figure}
\centerline{\epsfig{file=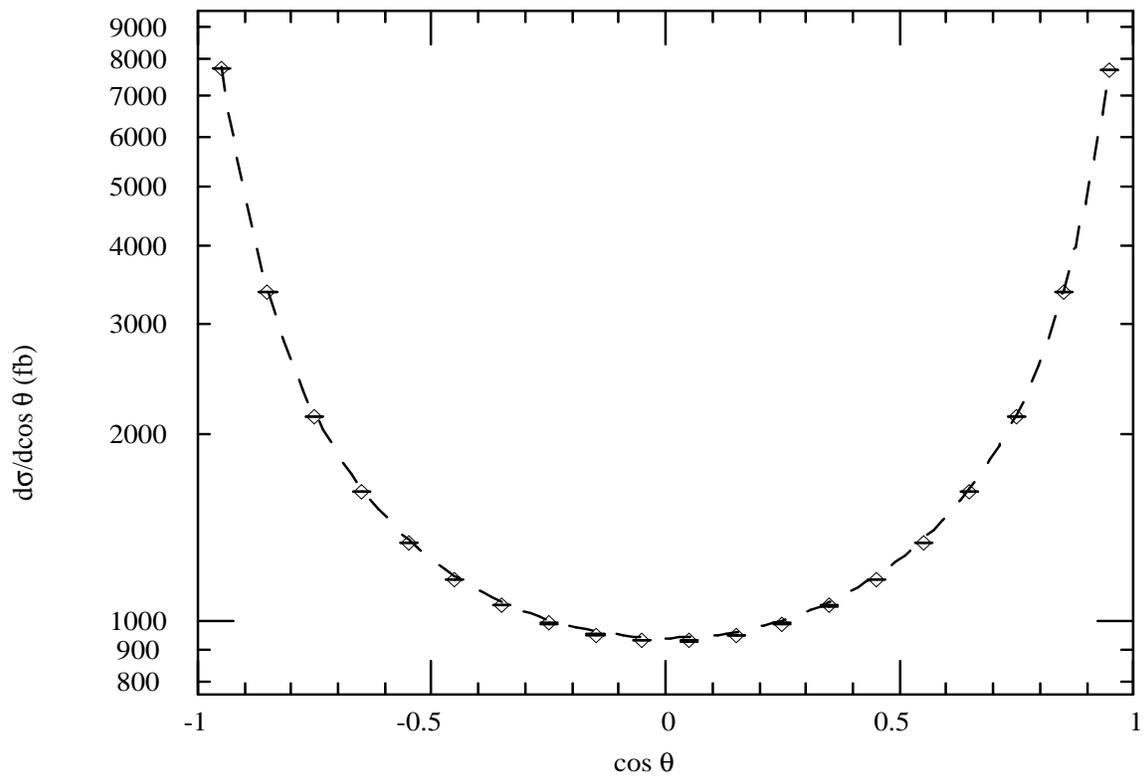,width=4in,angle=-90}}
\vspace{20pt}
\caption{$d\sigma/d\cos\theta$ vs. 
$\Lambda_{NC}$ 
for the 
pair production process, $\sqrt{s} = 500$ GeV, $\Lambda = 300$ GeV and 
$\alpha = 0$.  The dashed curve corresponds to the SM angular distribution 
and the points correspond to the NCQED angular distribution including 
1 standard deviation (statistical) error.}
\label{Fig3}
\end{figure}

\newpage
\begin{figure}
\begin{center}
\centerline{\epsfig{file=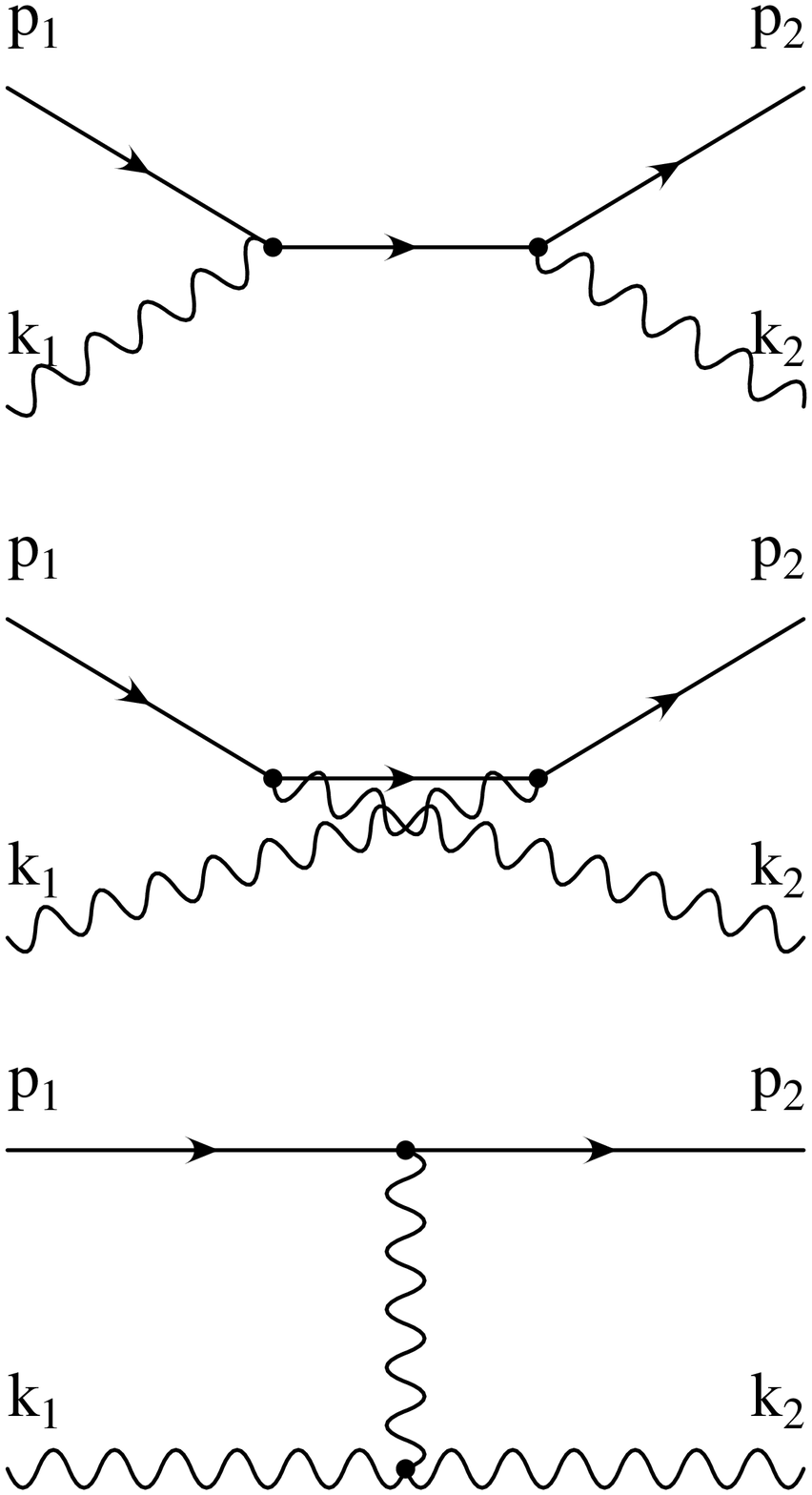,width=3.0in}}
\end{center}
\vspace{20pt}
\caption{The Feynman diagrams contributing to the process 
$e \gamma \to e \gamma$.}
\label{Fig5}
\end{figure}

\newpage
\begin{figure}
\centerline{\epsfig{file=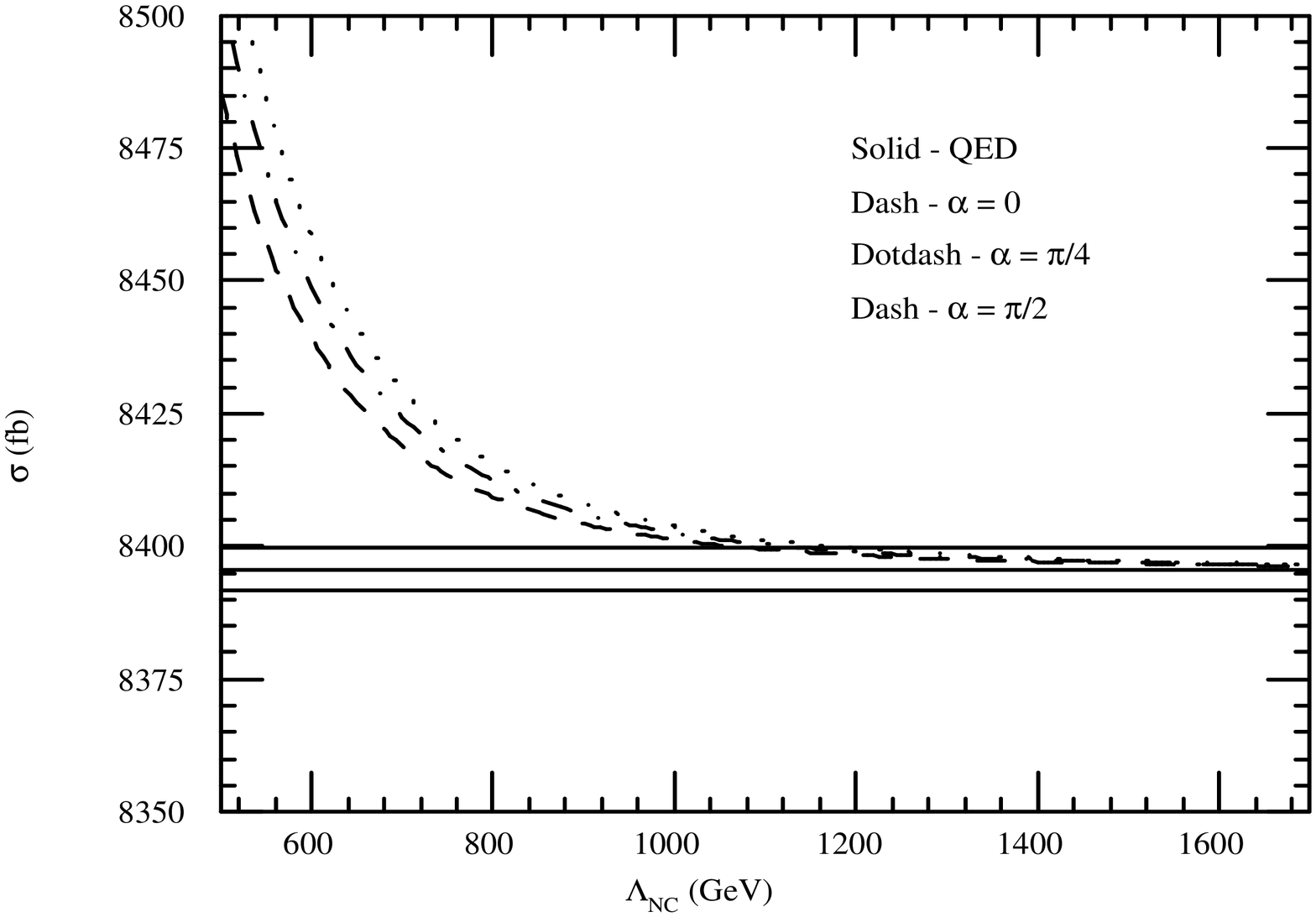,width=4.5in,angle=-90}}
\vspace{20pt}
\caption{$\sigma$ vs. $\Lambda_{NC}$ for the Compton scattering 
process with $\sqrt{s} = 500$ GeV
for $\gamma=0$.  The horizontal band represents the SM cross section 
$\pm$ 1 standard deviation (statistical) error.}
\label{Fig6}
\end{figure}

\newpage
\begin{figure}
\centerline{\epsfig{file=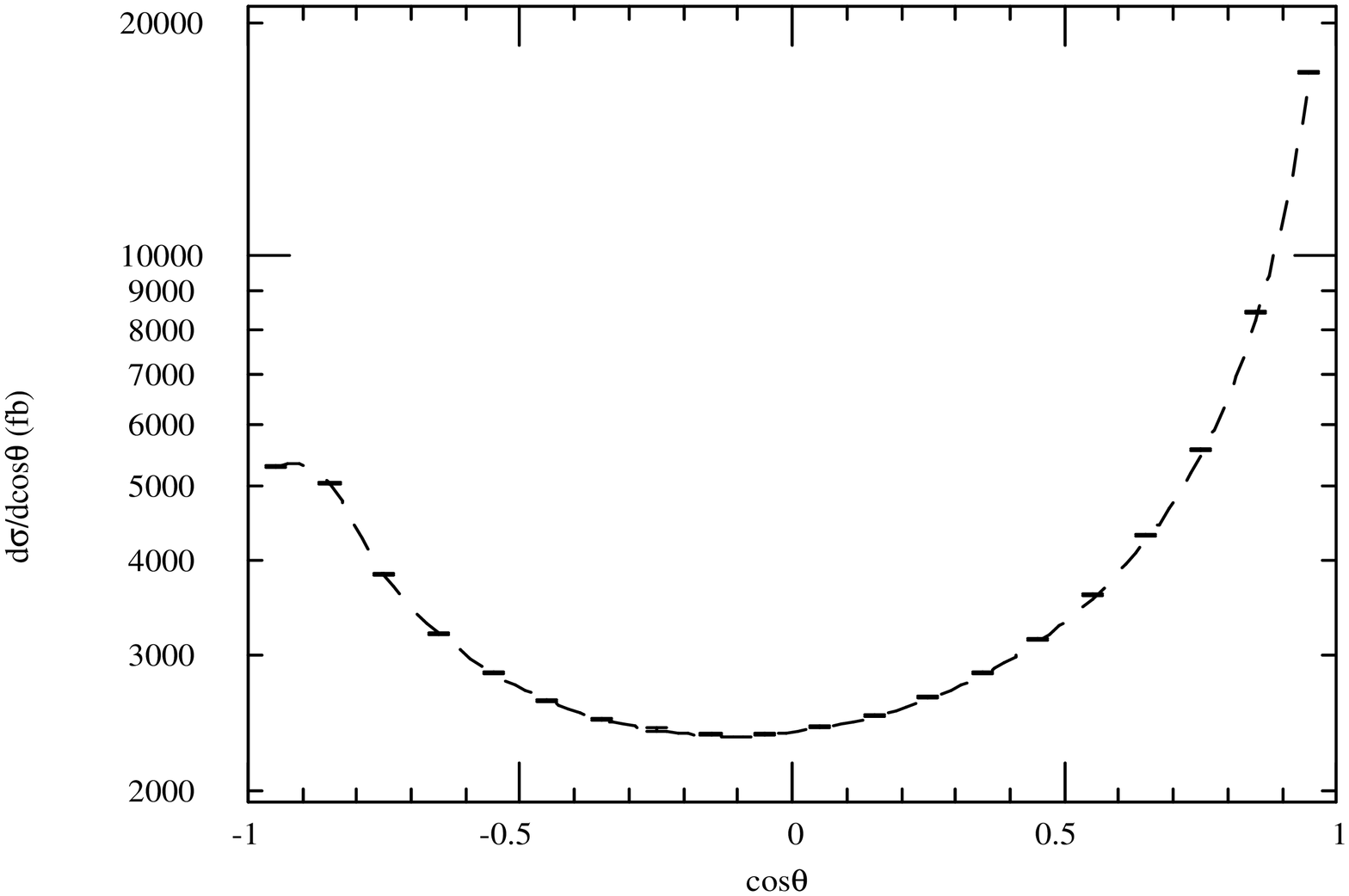,width=4in,angle=-90}}
\centerline{\epsfig{file=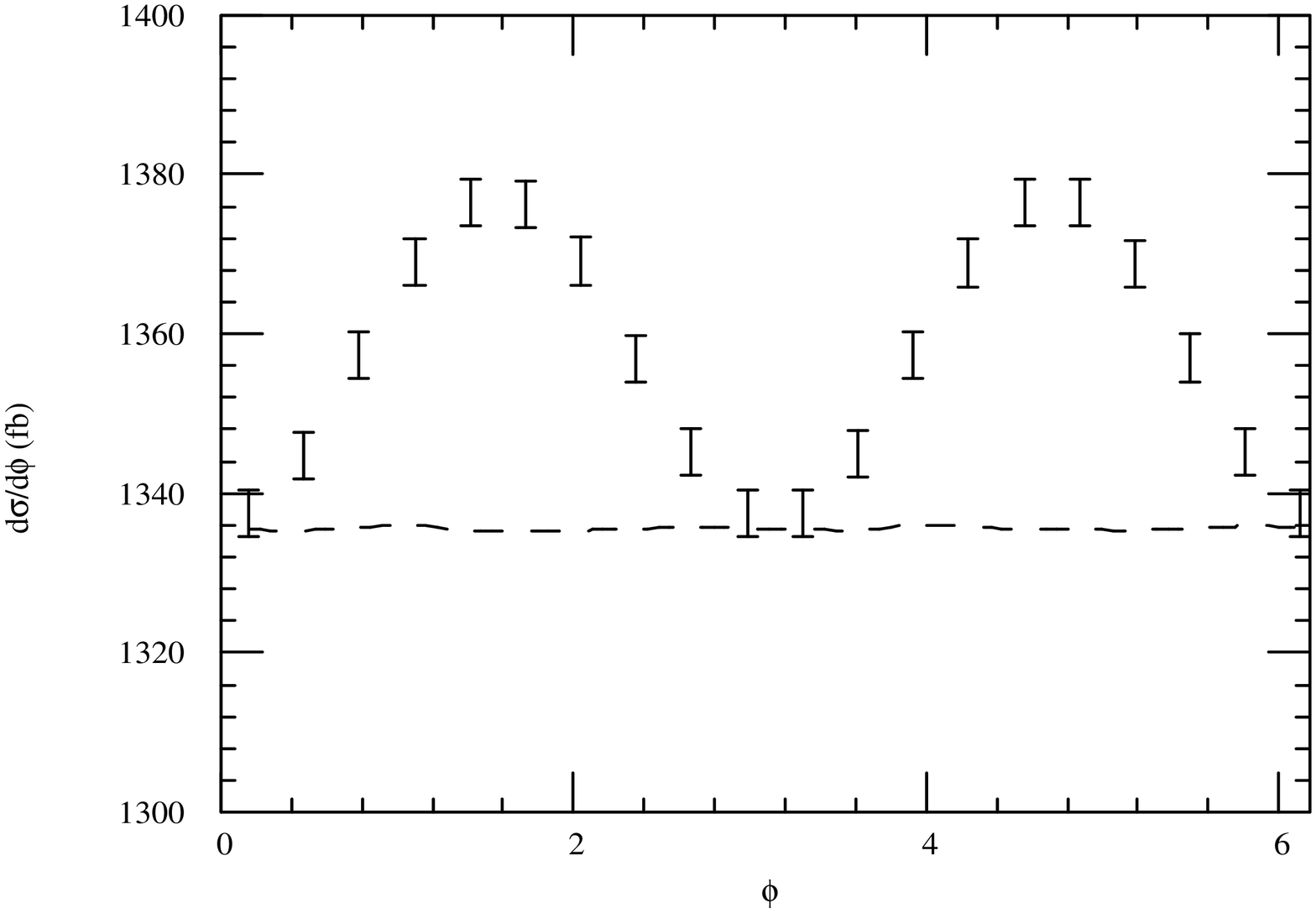,width=4in,angle=-90}}
\vspace{20pt}
\caption{(a) $d\sigma/d\cos\theta$  and (b) $d\sigma/d\phi$ for the 
Compton scattering process with $\sqrt{s} = 500$ GeV and for
 $\Lambda = 500$ GeV, $\alpha = \pi/2$ and $\gamma = 0$.  The dashed curve 
corresponds to the SM angular distribution 
and the points correspond to the NCQED angular distribution including 
1 standard deviation (statistical) error.}
\label{Fig7}
\end{figure}

\newpage
\begin{figure}
\centerline{\epsfig{file=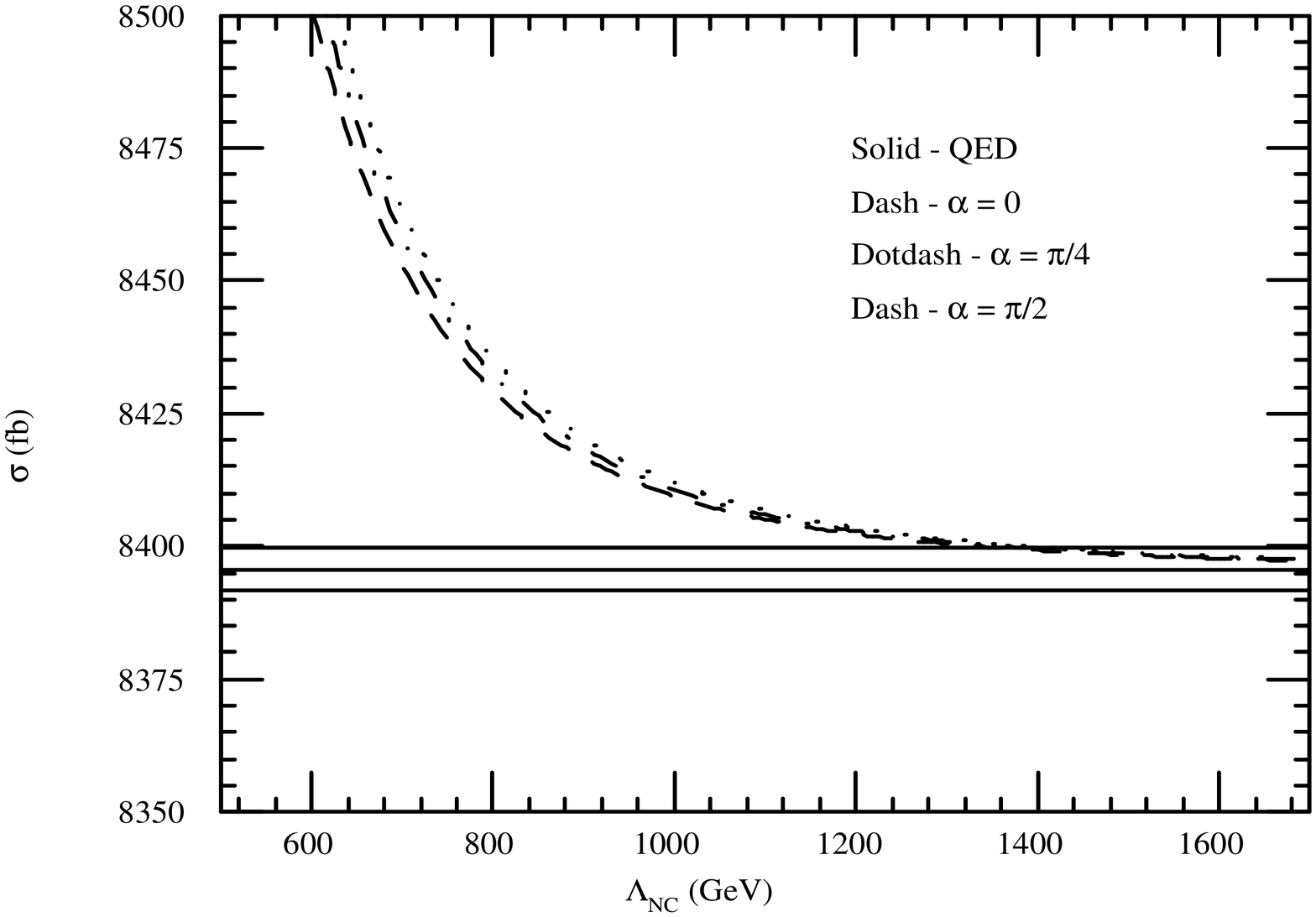,width=4.5in,angle=-90}}
\vspace{20pt}
\caption{$\sigma$ vs. $\Lambda_{NC}$ for the Compton scattering 
process with $\sqrt{s} = 500$ GeV for $\gamma=\pi/2$..  The horizontal band 
represents the SM cross section $\pm$ 1 standard deviation (statistical) 
error.}
\label{Fig9}
\end{figure}

\newpage
\begin{figure}
\centerline{\epsfig{file=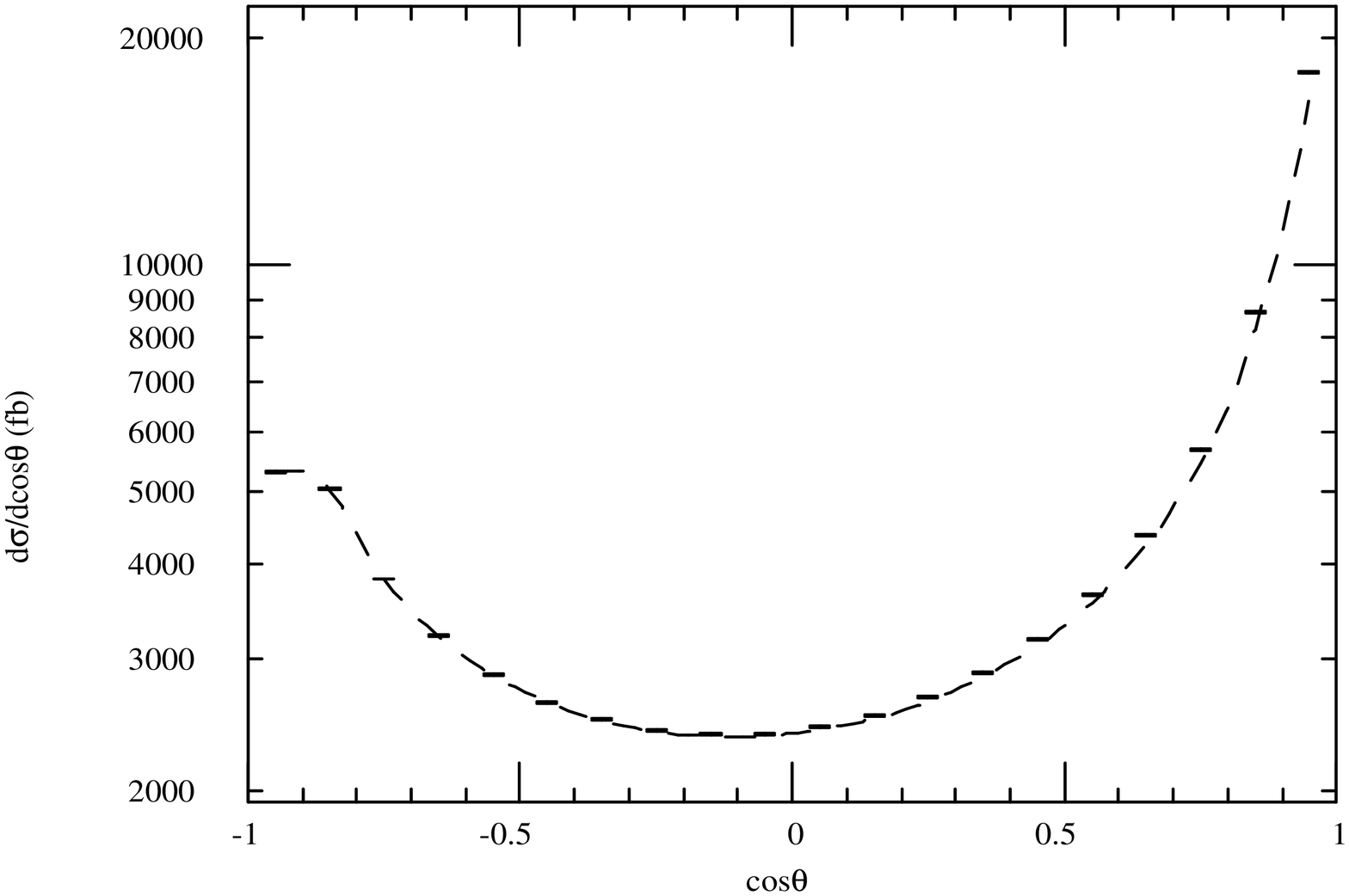,width=4in,angle=-90}}
\centerline{\epsfig{file=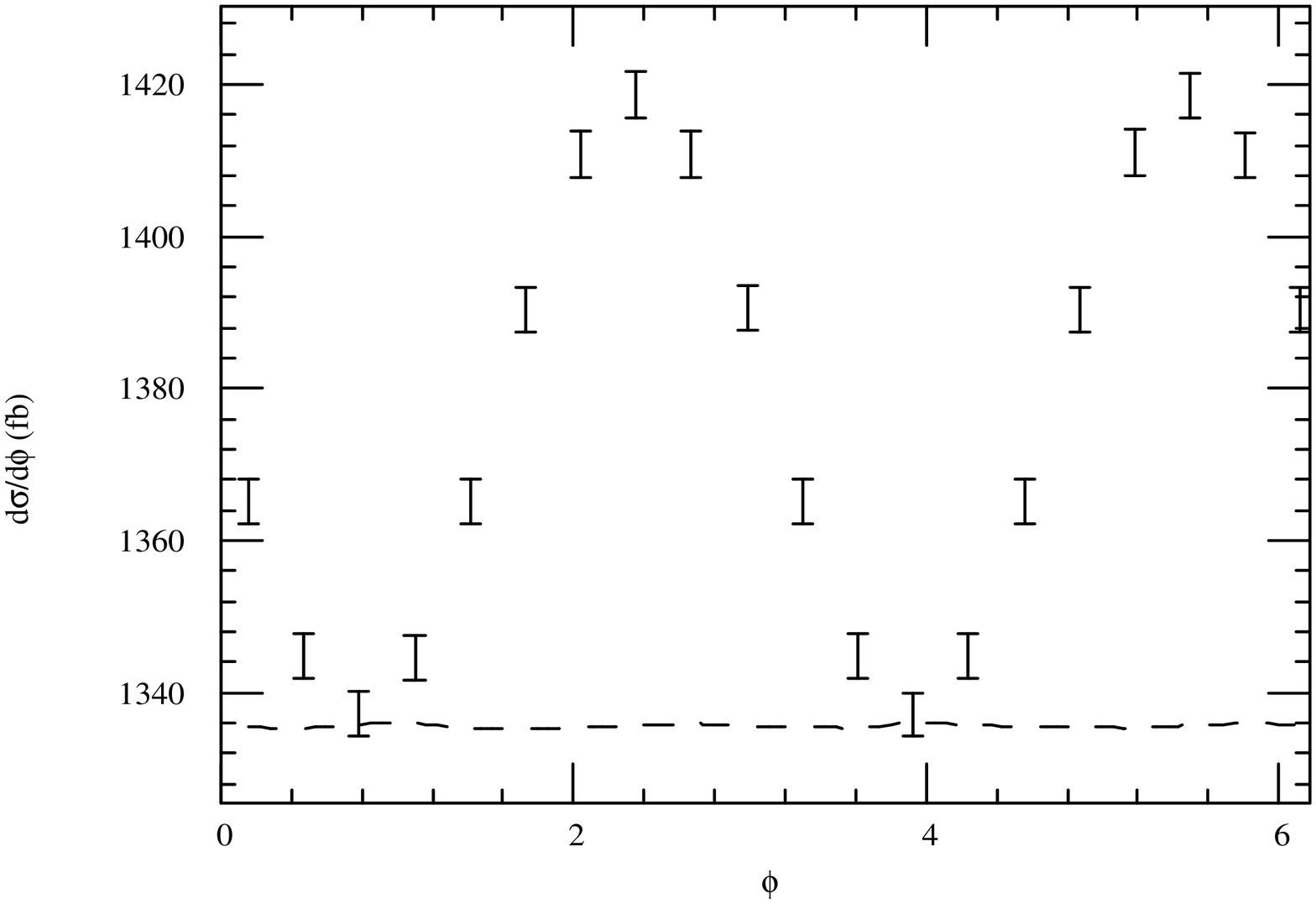,width=4in,angle=-90}}
\vspace{20pt}
\caption{(a) $d\sigma/d\cos\theta$  and (b) $d\sigma/d\phi$ for the Compton 
scattering process with $\sqrt{s} = 500$ GeV and for $\Lambda = 500$ GeV, 
$\alpha = \pi/2$ and  $\gamma = \pi/2$.  The dashed curve corresponds to the 
SM angular distribution and the points correspond to the NCQED angular 
distribution including 1 standard deviation (statistical) error.}
\label{Fig10}
\end{figure}

\clearpage

\newpage
\begin{table}[p]
\caption{95\% C.L.\ exclusion limits, in GeV, for the pair production process 
at a $\gamma \gamma$ collider. 
Results are presented for $\sqrt{s}=0.5$, 0.8, 1.0, 1.5, 3.0 and 5.0 TeV and 
for two values of $\alpha$, 0 and $\pi/4$.  The $\delta^{stat}$ column 
is with no systematic error included and the column labelled 
$\delta^{stat}+\delta^{sys}$ includes a 2\% systematic error.}
\label{limitspp}
\vspace{0.4cm}
\begin{center}
\begin{tabular}{lrrrr}
$\sqrt{s}$ & \multicolumn{2}{c}{$\alpha = 0$} & 
	\multicolumn{2}{c}{$\alpha= \pi/4$ } \\
(TeV)      &  $\delta^{stat}$ & $\delta^{stat}+\delta^{sys}$  
		&  $\delta^{stat}$ & $\delta^{stat}+\delta^{sys}$  \\
\hline
0.5        &  535 & 260 &  445 & 220  \\
0.8        &  740 & 400  &  620 & 335     \\
1.0        &  860 & 485   &  725 & 405      \\
1.5        & 1145 & 700   &  965 & 590      \\
3.0        & 1880 & 1320  & 1580 & 1110     \\
5.0        & 2700 & 2090  & 2270 & 1760     \\
\end{tabular}
\end{center}
\end{table}

\newpage
\begin{table}[p]
\caption{95\% C.L.\ exclusion limits, in GeV, for the Compton scattering 
process.  Results are presented for $\sqrt{s}=0.5$, 0.8, 1.0, 1.5, 3.0 and 5.0 
TeV and for $\gamma = 0$ and  $\gamma = \pi/2$
and for three values of $\alpha$, 0, $\pi/4$ and 
$\pi/2$.  
The $\delta^{stat}$ column 
is with no systematic error included and the column labelled 
$\delta^{stat}+\delta^{sys}$ includes a 2\% systematic error.
}
\label{limitscs0}
\vspace{0.4cm}
\begin{center}
\begin{tabular}{lrrrrrr}
$\sqrt{s}$ (TeV) & \multicolumn{6}{c}{$\gamma = 0$} \\
\hline
 	& \multicolumn{2}{c}{$\alpha = 0$}  
	&	\multicolumn{2}{c}{$\alpha= \pi/4$ } 
	&	\multicolumn{2}{c}{$\alpha= \pi/2$ } \\
\hline
    	&  $\delta^{stat}$ & $\delta^{stat}+\delta^{sys}$  
	&  $\delta^{stat}$ & $\delta^{stat}+\delta^{sys}$  
	&  $\delta^{stat}$ & $\delta^{stat}+\delta^{sys}$\\
\hline
0.5        &  925 & 545   & 1020  & 585     & 1100  & 600    \\
0.8        & 1325 & 875   & 1455  & 935     & 1565  & 960      \\
1.0        & 1565 & 1090  & 1720  & 1165    & 1850  & 1200     \\
1.5        & 2125 & 1620  & 2330  & 1740    & 2505  & 1785     \\
3.0        & 3575 & 3110  & 3920  & 3375    & 4220  & 3460     \\
5.0        & 5240 & 4880  & 5745  & 5325    & 6185  & 5465     \\
\hline
$\sqrt{s}$ (TeV) & \multicolumn{6}{c}{$\gamma = \pi/2$} \\
\hline
0.5        & 1215 & 700   & 1245 & 715      & 1305 & 720       \\
0.8        & 1730 & 1115  & 1780 & 1135     & 1860 & 1140      \\
1.0        & 2045 & 1390  & 2100 & 1415     & 2200 & 1425      \\
1.5        & 2770 & 2070  & 2845 & 2110     & 2980 & 2125      \\
3.0        & 4660 & 4010  & 4785 & 4085     & 5015 & 4115      \\
5.0        & 6840 & 6335  & 7020 & 6460     & 7360 & 6500      \\
\end{tabular}
\end{center}
\end{table}

\end{document}